\documentclass{article}
\usepackage{cite}
\usepackage{graphicx}
\usepackage{dcolumn}

\begin{document}

\date{}
\title{On the effective secular equation}
\author{Francisco M. Fern\'{a}ndez \\
INIFTA, Divisi\'on Qu\'imica Te\'orica\\
Blvd. 113 (S/N), Sucursal 4, Casilla de Correo 16, 1900 La Plata, Argentina}
\maketitle

\begin{abstract}
We show that the effective secular equation proposed several years ago is
suitable for estimating the location of the exceptional points of eigenvalue
equations. As an illustrative example we choose the well known Mathieu
equation.
\end{abstract}

\section{Introduction}

\label{sec:intro}

Several years ago, Fried and Ezra\cite{FE89} developed a method for
improving the calculation of eigenvalues by means of their perturbation
series. The resummation method consists of the reconstruction of an
effective secular equation (ESE) that yields better results than the
original perturbation series. They applied the approach to the Barbanis
Hamiltonian.

Somewhat later, Zheng\cite{Z22} presented essentially the same approach and
applied it to a trivial tridiagonal $3\times 3$ matrix representation of a
toy Hamiltonian operator. It has been shown that both approaches yield the
same result on the just mentioned oversimplified model\cite{F22}. However,
Zheng\cite{Z22b} put forward some unconvincing arguments with the purpose of
showing that there are some relevant differences between both methods. In
particular, Zheng focused on the fact that the ESE yielded complex
eigenvalues for the Barbanis Hamiltonian\cite{FE89}. Clearly, Zheng
overlooked the fact that the anharmonic oscillator chosen by Fried and Ezra%
\cite{FE89} does not exhibit bound states and, consequently, the appearance
of complex eigenvalues is not surprising. What is more, Zheng\cite{Z22b} did
not attempt to apply his approach to the Barbanis Hamiltonian or to compare
his results with those derived by Fern\'{a}ndez\cite{F22} using the ESE
method.

In a later paper, Zheng\cite{Z22c} proposed a most unclear
strategy to improve the ESE approach of Fried and Ezra to which he
refers, quite pejoratively, as \textit{cockamamie}. Zheng never
attempted to test his new approach on a demanding problem and
after three papers\cite{Z22,Z22b,Z22c} he did not go beyond an
extremely simple toy model given by a $3\times 3$ tridiagonal
matrix representation of an Hermitian operator. Zheng's supposed
improvement of ESE\cite{Z22c} is based on the determinant of $\frac{E-H_{%
\mathrm{eff}}(\lambda )}{E-H_{0}^{P}}$. Since the operators $H_{\mathrm{eff}%
}(\lambda )$ and $H_{0}^{P}$ are not expected to commute it is not clear
whether he means either $\left( E-H_{0}^{P}\right) ^{-1}\left[ E-H_{\mathrm{%
eff}}(\lambda )\right] $ or $\left[ E-H_{\mathrm{eff}}(\lambda )\right]
\left( E-H_{0}^{P}\right) ^{-1}$.

As mentioned above, Fried and Ezra\cite{FE89} applied ESE to the Barbanis
Hamiltonian. Unfortunately, this nontrivial problem is not the most suitable
one for testing the approach because the perturbation series exhibit zero
radius of convergence. On the other hand, the perturbation series for the
eigenvalues of the toy problem chosen by Zheng\cite{Z22} exhibit finite
radii of convergence and one can locate the exceptional point closest to
origin by means of ESE\cite{F22}. In this paper we apply ESE to a nontrivial
problem in which the perturbation expansions exhibit finite radii of
convergence, the Mathieu equation.

\section{The effective secular equation}

\label{sec:ESE}

In this section we develop ESE following the lines of our earlier paper\cite
{F22} based on Fried and Ezra's one\cite{FE89}. Suppose that we want to
obtain the solutions to the Schr\"{o}dinger equation $H\psi _{n}=E_{n}\psi
_{n}$, $n=1,2,\ldots $, where $H=H_{0}+\lambda H_{I}$. To this end, we apply
perturbation theory and obtain partial sums of the form
\begin{equation}
E_{n}^{[K]}=\sum_{j=0}^{K}E_{n,j}\lambda ^{j}.  \label{eq:E_n^[K]}
\end{equation}
The ESE is given by
\begin{equation}
\left\{ \prod_{n=1}^{N}\left[ W-E_{n}^{[K]}(\lambda )\right] \right\}
^{[K]}=W^{N}+\sum_{j=1}^{N}p_{j}(\lambda )W^{N-j},  \label{eq:F(W,lamb)}
\end{equation}
where $\left\{ ...\right\} ^{[K]}$ means that we remove any term with $%
\lambda ^{j}$ if $j>K$. The reason is that the accuracy of the results
cannot be greater than $\mathcal{O}\left( \lambda ^{K}\right) $ determined
by the partial sums (\ref{eq:E_n^[K]}). The eigenvalues $E_{n}$ used in this
reconstruction are related to the so-called model space, which is finite,
and we leave aside the complement space that is not necessarily so. The
states in the model space are strongly coupled among themselves and weakly
coupled to the states in the complement space. The roots $W_{n}(\lambda )$, $%
n=1,2,\ldots ,N$, of the polynomial (\ref{eq:F(W,lamb)}) are expected to be
better approximations to the eigenvalues $E_{n}(\lambda )$, $n=1,2,\ldots ,N$%
, that the partial sums $E_{n}^{[K]}(\lambda )$ used in the construction of
the ESE.

\section{The Mathieu equation}

\label{sec:Mathieu}

As an example we choose the Hermitian operator
\begin{equation}
H=-d^{2}/dx^{2}+2\lambda \cos (2x),  \label{eq:H_Mathieu}
\end{equation}
so that $H\psi =E\psi $ yields the well known Mathieu equation. We may
consider solutions of period $\pi $ ($\psi (x+\pi )=\psi (x)$) and $2\pi $ ($%
\psi (x+2\pi )=\psi (x)$) and each class can be separated into sub classes
with even ($\psi (-x)=\psi (x)$) and odd ($\psi (-x)=-\psi (x)$)
eigenfunctions $\psi (x)$. In a recent paper Amore and Fern\'{a}ndez\cite
{AF21} obtained some of the exceptional points for the Mathieu eigenvalue
equation quite accurately. Here, we simply try to estimate the exceptional
point $\lambda _{p}$ closest to origin by means of ESE.

One can easily obtain the perturbation expansions for the eigenvalues of $H$
by means of well known suitable approaches\cite{F01}. For example, the
perturbation series for the two lowest eigenvalues of the $2\pi $ even
subspace are
\begin{eqnarray}
E_{1} &=&1+\lambda -\frac{\lambda ^{2}}{8}-\frac{\lambda ^{3}}{64}-\frac{%
\lambda ^{4}}{1536}+O\left( \lambda ^{5}\right) ,  \nonumber \\
E_{2} &=&9+\frac{\lambda ^{2}}{16}+\frac{\lambda ^{3}}{64}+\frac{13\lambda
^{4}}{20480}+O\left( \lambda ^{5}\right) ,  \label{eq:E_n_PT}
\end{eqnarray}
and we can obtain as many perturbation corrections as desired.
Straightforward application of ESE as outlined in section~\ref{sec:ESE} with
$N=2$ and increasing values of $K$ enables us to obtain
\begin{equation}
\begin{array}{ccc}
K & \left| \lambda _{p}\right|  & \lambda _{p} \\
10 & 3.769959083 & 1.931394919\pm 3.237638825i \\
11 & 3.769957228 & 1.931392571\pm 3.237638065i \\
12 & 3.769957375 & 1.931392656\pm 3.237638186i \\
13 & 3.769957431 & 1.931392443\pm 3.237638378i
\end{array}
\label{eq:lam_p_Mathieu}
\end{equation}
It is understood that $E_{1}(\lambda )$ and $E_{2}(\lambda )$ coalesce at
the exceptional point $\lambda _{p}$ and the radius of convergence of the
series (\ref{eq:E_n_PT}) is $\left| \lambda _{p}\right| $. On the other
hand, the application of the discriminant to the secular equation for $H$ in
the $2\pi $ even subspace yielded $\left| \lambda _{p}\right| =3.769957494$
that is accurate to the last digit\cite{AF21} . We appreciate that ESE
enables us to accurately estimate the exceptional point closest to origin
from the perturbation series as suggested by earlier calculations on the toy
model\cite{Z22,F22}.

Zheng\cite{Z22c} stated that ``This reconstruction is also inefficient. Even
to order of $\lambda ^{2}$, one has to perform a second-order perturbation
calculation for all the $E_{n}^{P}(\lambda )$, and some of these
calculations actually are redundant since we have proved that terms such as
those containing $\frac{1}{\epsilon _{i}^{P}-\epsilon _{n}^{P}}$ must offset
each other in the final effective secular equation.'' We must confess that
we do not understand this statement. In our example above, we used
perturbations expansions of order $\lambda ^{13}$ without difficulty and the
results shown in (\ref{eq:lam_p_Mathieu}) exhibit a remarkable rate of
convergence.

\section{Conclusions}

\label{sec:conclusions}

It is clear that Zheng\cite{Z22b} failed to show that his approach is
different from the ESE of Fried and Ezra\cite{FE89}. One reason for such
statement is that he did not apply his approach to the Barbanis Hamiltonian.
In addition to it, the accurate Fern\'{a}ndez's results for the exceptional
point of the toy problem\cite{F22} agree with two of the three results shown
by Zheng\cite{Z22}. The discrepancy for the order $\lambda ^{2}$ appears to
be a misprint in Zheng's paper.

Zheng's improvement\cite{Z22c} of the ESE approach of Fried and Ezra\cite
{FE89} is based on a wrong expression that does not take into account that
the operators involved do not commute. Besides, Zheng did not show any
calculation to convince the readers about the advantages of his proposal.

In section~\ref{sec:Mathieu} we calculated one of the exceptional points of
the Mathieu equation by means of the ESE approach of Fried and Ezra\cite
{FE89}. Although this problem is relatively simple, it is not trivial. On
the other hand, Zheng\cite{Z22,Z22b,Z22c} did not go beyond the tridiagonal $%
3\times 3$ matrix representation of a toy Hermitian Hamiltonian.

\end{document}